\documentclass[aps,eqsecnum,nofootinbib,superscriptaddress,showpacs]{revtex4}
\usepackage[dvips]{graphicx}  
\usepackage{amsmath,amsfonts}
\usepackage{color}  
\def\be{\begin{eqnarray}}
\def\ee{\end{eqnarray}}
\def\beq{\begin{equation}}
\def\eeq{\end{equation}}

\def\g{\gamma}

\def\({\left (}
\def\){\right )}

\newtheorem{theorem}{Theorem}[section]

\newtheorem{proposition}[theorem]{Proposition}
\newtheorem{corollary}[theorem]{Corollary}
\newtheorem{definition}{Definition}[section]

\newcommand{\qed}{\nobreak \ifvmode \relax \else
      \ifdim\lastskip<1.5em \hskip-\lastskip
      \hskip1.5em plus0em minus0.5em \fi \nobreak
      \vrule height0.75em width0.5em depth0.25em\fi}
\newcommand{\qcd}{\begin{flushright} $\Box$ \end{flushright}}
\begin{document}

\title{
On the Gannon-Lee Singularity Theorem in Higher Dimensions
}

\author{I.P. Costa e Silva}
\email{pontual.ivan@gmail.com}
\affiliation{Department of Mathematics,\\ 
Universidade Federal de Santa Catarina \\88.040-900 Florian\'{o}polis-SC, Brasil}

\date{\today}

\begin{abstract}
The Gannon-Lee singularity theorems give well-known restrictions on the spatial topology of singularity-free (i.e., nonspacelike geodesically complete), globally hyperbolic spacetimes. In this paper, we revisit these classic results in the light of recent developments, especially the failure in higher dimensions of a celebrated theorem by Hawking on the topology of black hole horizons. The global hyperbolicity requirement is weakened, and we expand the scope of the main results to allow for the richer variety of spatial topologies which are likely to occur in higher-dimensional spacetimes.         
              
\end{abstract}
\maketitle

\section{Introduction}\label{sec:intro}

The interplay between the causal and topological aspects of spacetimes is a topic of permanent interest both in physics and mathematics. Of course, such a relationship between different ``levels of structure'' on a spacetime (viewed as a purely geometrical object) is the stock-in-trade of Lorentzian Geometry. On the physical side, a natural question is whether constraints on the topology imposed by the causal structure can have physical appeal and yet account for the fact that one does not apparently observe any non-trivial features like handles or holes in spacetime.

The singularity theorems proved independently by Gannon \cite{gannon1,gannon2} and Lee \cite{lee} in 1975/6, were among the earliest fruitful attempts to address this question. As an example, consider the following statement of one of Gannon's theorems \cite{gannon1}: 

\begin{theorem}[Gannon 1975] 
\label{standardgannon}
Let $M^n$ be a spacetime which satisfies the null energy condition [i.e., $\mbox{Ric} (v,v) \geq 0$ for any null vector $v \in TM^n$] , and admits a smooth, spacelike Cauchy hypersurface $\Sigma \subseteq M$ which is regular near infinity. If $\Sigma$ is non-simply connected, then $M^n$ is null geodesically incomplete.
\end{theorem} 

(Here, $n$ denotes the dimension of spacetime: although the arguments given in \cite{gannon1,gannon2,lee} are developed in $n=4$, they naturally go through for $n>4$ as well.) The distinguishing points in Gannon's theorem which contrast with the classic Penrose-Hawking singularity theorems \cite{penrose,HE} are that (1) the existence of a {\em regular near infinity} Cauchy hypersurface $\Sigma$ is assumed (in Lee's version of this theorem, a different but related hypothesis is made), and (2) $\Sigma$ cannot have certain non-trivial topological features such as handles if spacetime is non-singular. Leaving (1) aside for the moment, a natural, intuitive picture that emerges from the conclusion (2) is that non-simply connected sections of spacetime would collapse into a singularity under reasonable physical conditions. This scenario is reinforced by the {\em topological censorship theorems} \cite{FSW,galloway1,galloway2,galloway3,galloway4}, which under conditions similar to some of the ones we have in Theorem \ref{standardgannon} imply that these non-simply connected sections and the ensuing singularity must be hidden from active scrutiny (in a suitable sense) by a distant observer. 

The condition that $\Sigma$ be regular near infinity is a kind of asymptotic flatness requirement whose precise phrasing need not concern us just now. Suffice it to say that it implies in particular that $\Sigma$ can be divided in two disjoint open regions $\Sigma_{+}$ and $\Sigma_{-}$ separated by a common boundary $S$, where $S$ is diffeomorphic to a (n-2)-sphere, $\Sigma_{+}$ is homeomorphic to $S \times \mathbb{R}$ and represents the ``outside" asymptotic region, and $K=\Sigma_{-} \cup S$ is compact, with all eventual topological complexities lumped into the interior of $K$. If topological censorship holds, one is naturally led to think of $S$ as a spacelike section of an event horizon hiding handles and the like from view. Thus presented, regularity near infinity would seem harmless enough. 

However, the assumption that $S$ has the topology of a (n-2)-sphere ceases to be natural in spacetime dimensions higher than four. Indeed, a well-known theorem by Hawking \cite{hawking,HE} establishes that any connected component of cross sections of the event horizon in a 4-dimensional, asymptotically flat stationary black hole spacetime obeying the dominant energy condition is homeomorphic to a 2-sphere. This is a fundamental result in black hole theory, but a naive analog of this theorem is false in dimensions higher than four. A now famous counterexample is the 5-dimensional stationary vacuum black hole spacetime discovered by Emparan and Reall, with horizon topology $S^2 \times S^1$ \cite{ER1,ER2}. A sensible generalization was given by Schoen and Galloway \cite{SG,galloway5}: under the hypotheses of Hawking's theorem, the cross sections of black hole event horizons have {\em positive Yamabe type}, which places a number of well-known topological restrictions \cite{SY,GL} in any dimension, and in four dimensions recovers Hawking's result (see also \cite{helfgott} for a different set of restrictions). Actually, the Schoen-Galloway theorem is a general result about the topology of certain marginally outer trapped surfaces (MOTS). Our separation surface $S$ could in principle be one such MOTS, and assuming it is a sphere from the outset would then seem unnatural. 
Moreover, even in four spacetime dimensions, Hawking's result is known to fail if the dominant energy condition \cite{GH} or asymptotic flatness \cite{galloway6} do not hold.   

The purpose of this paper is to provide a generalization of the Gannon-Lee theorem more suitable to a higher-dimensional context. The rest of it is organized as follows. We first give some preliminary definitions in Section \ref{preliminaries} to set the conventions, general assumptions and notation, and state the main Theorem. The particular causality condition we use, although well-known, does not appear very frequently in singularity theorems, so we give more details on some of its aspects in Section \ref{causalsimplicity}. The proofs of the main theorem together with some auxiliary results are deferred to Section \ref{detailedproofs}. We end with some general remarks on the main result. 

We should mention that Gannon and Lee each proved a number of closely related results in their respective works, although their statements of similar theorems are somewhat distinct in appearance. We believe that Gannon's theorem stated above neatly synthesizes the main thrust of their ideas, and accordingly this is the template for our main theorem. But since Lee rediscovered essentially the same results at about the same time, we shall, in what follows, refer to this theorem as `the Gannon-Lee (singularity) theorem'.

\section{Preliminaries \& Main Theorem}\label{preliminaries}

 In what follows, we fix a {\em spacetime}, i.e, an $n$-dimensional, second-countable, connected, Hausdorff, $C^{\infty}$ time-oriented Lorentz (signature $(-,+, \ldots,+)$) manifold $M$ endowed with a smooth metric tensor $g$ and with $n \geq 3$. We assume that the reader is familiar with the basic definitions and results of global Lorentzian geometry and causal theory of spacetimes as found in the core references \cite{HE,wald,oneill,BE}, and in particular the Hawking-Penrose singularity theorems. We shall also assume familiarity with standard facts about covering manifolds and fundamental groups which can be found, for example, in Ref. \cite{massey}. All submanifolds of $M$ are embedded, and their topology is the induced topology. Finally, we follow the convention that causal vectors are always nonzero.

 Fix a smooth, connected, spacelike partial Cauchy hypersurface (i.e, a submanifold of codimension one) $\Sigma \subset M$ \footnote {Recall that a {\em partial Cauchy hypersurface} is by definition an acausal edgeless subset of a spacetime, which means in particular that it is a topological (i.e. $C^0$) hypersurface \cite{oneill}. In this paper, however, we always deal with smooth hypersurfaces.}, and a smooth, connected, compact spacelike submanifold of codimension two $S \subset M$. 

 Suppose $S$ {\em separates} $\Sigma$, i.e., $S \subset \Sigma$ and $\Sigma \setminus S$ is not connected. This means, in particular, that $\Sigma \setminus S$ is a disjoint union $\Sigma_{+}\dot{\cup} \Sigma_{-}$ of open submanifolds of $\Sigma$ having $S$ as a common boundary. We shall loosely call $\Sigma_{+}$ [resp. $\Sigma_{-}$] the {\em outside} [resp. {\em inside}] of $S$ in $\Sigma$. (In most interesting examples there is a natural choice for these.) It also means that there are unique unit spacelike vector fields $N_{\pm}$ on $S$ normal to $S$ in $\Sigma$, such that $N_{+}$ [resp. $N_{-}$] is outward-pointing [resp. inward-pointing]. Thus, it follows that $S$ is {\em two-sided} in $\Sigma$ \cite{oneill}, i.e., there exists a smooth embedding $F:S \times (-1,1) \rightarrow \Sigma$ such that $F(p,0)=p$, for all $p \in S$, and ${\cal U}_F^{\pm} \subseteq \Sigma_{\pm}$, where ${\cal U}_F^{+} := F(S \times (0,1))$ and ${\cal U}_F^{-} := F(S \times (-1,0))$. Therefore, the (open) image $F(S \times (-1,1)) \equiv {\cal U}_F^{+} \cup S \cup {\cal U}_F^{-} \subseteq \Sigma$ of such a map is a {\em collar } of $S$ and can be foliated by a one-parameter family of diffeomorphic copies of $S$. We call any such embedding $F$ a {\em local deformation} of $S$ (in $\Sigma$).    

Let $U$ be the unique timelike, future-directed, unit normal vector field on $\Sigma$. Then $K_{\pm} := U|_{S}+N_{\pm}$ are future-directed null vector fields on $S$ normal to $S$ in $M$. The {\em outward }[resp. {\em inward}] {\em null convergence} of $S$ in $M$ is the smooth function $k_{+}: S \rightarrow \mathbb{R}$ [resp. $k_{-}: S \rightarrow \mathbb{R}$] given by 
\begin{equation}
\label{maineq1}
k_{+}(p) = \langle H_p, K_{+}(p)\rangle _{p} \mbox{  [resp. $k_{-}(p) = \langle H_p, K_{-}(p) \rangle_p $]},
\end{equation}
for each $p \in S$, where $H_p$ denotes the mean curvature vector of $S$ in $M$ at $p$ \cite{oneill}, and we denote $g$ as $\langle \, , \, \rangle$ here and hereafter, if there is no risk of confusion. \footnote{Note that we have $k_{\pm} = - \theta_{\pm}$, where $\theta_{\pm}$ are the null expansion scalars defined, e.g., in Ref. \cite{SG}.} 

Physically, these functions measure the convergence of light rays emanating from $S$. If $S$ is a round sphere in a Euclidean slice of Minkowski spacetime, with the obvious choices of inside and outside, we have $k_{+}<0$ and $k_{-}>0$. One also expects this to be the case if $S$ is a ``large"  sphere in an asymptotically flat spacetime. But in a region of strong gravity one expects instead that we have both $k_{+}>0$ and $k_{-}>0$, in which case $S$ is a closed (i.e., compact and without boundary) trapped surface.

 We say that a smooth future-directed timelike vector field $X:M \rightarrow TM$ is a {\em piercing} of $\Sigma$ (or {\em pierces} $\Sigma$) if every maximally extended integral curve of $X$ intersects $\Sigma$ exactly once. In physical terms, one may think of the integral curves of a piercing as worldlines of members of a family of observers who ``witness"  the ``whole universe at a certain instant of common time" described by $\Sigma$. Although in what follows the existence of a piercing will be regarded as a technical tool in lieu of global hyperbolicity in the proof of our main theorem, the above interpretation will hopefully convince the reader that it is a rather harmless assumption from a physical standpoint. Moreover, it is not difficult to check that a piercing does exist for suitable partial Cauchy hypersurfaces in basic solutions like Minkowski, Kerr-Newman and FRW spacetimes.  

 Of course, a piercing of $\Sigma$ may not exist for general spacetimes. If $M$ is compact for example, the integral lines of any vector field will tend to recur by the Poincar\'{e} Recurrence Theorem (see, e.g., \cite{AM}, pg. 208), and might thus intersect $\Sigma$ more than once. On the other hand, if $(M,g)$ is globally hyperbolic and $\Sigma$ is a Cauchy hypersurface, then {\em every} smooth future-directed timelike vector field in $M$ pierces $\Sigma$. {\em However, the existence of a piercing for $\Sigma$ is strictly weaker than the requirement that $\Sigma$ be Cauchy}. For example, consider 4-dimensional anti-de Sitter spacetime, taken to be $\mathbb{R}^4$ with the metric given by the line element (see, e.g., \cite{HE}, pg. 131)
\begin{equation}
\label{AdS}
ds^2 = -\cosh ^2rdt^2 + dr^2 + \sinh ^2 r \left(d\theta ^2 + \sin ^2 \theta d \phi ^2 \right),
\end{equation}                
where the coordinate ranges are $-\infty < t < \infty$, $r >0$, $0< \theta <\pi$, and $0 < \phi < 2\pi$. This spacetime is not globally hyperbolic, but each hypersurface $t = \mbox{const.}$, although not Cauchy, is pierced by the vector field $\frac{\partial}{\partial t}$. 

 Using the terminology above, we shall adopt the following useful definition:
 
 \begin{definition}
 \label{asympticallyregular}
 A smooth, connected, spacelike partial Cauchy hypersurface $\Sigma \subset M$ is {\em asymptotically regular} if there exists a smooth, connected, compact submanifold $S \subset \Sigma$ of dimension $n-2$ such that
\begin{itemize}
\item[i)] $S$ separates $\Sigma$, and $\overline{{\Sigma}}_{+} \equiv S \cup \Sigma_{+}$ is non-compact;
\item[ii)] The map $h_{\#}: \pi_1(S) \rightarrow \pi_1(\overline{\Sigma}_{+})$ induced by the inclusion $h: S \hookrightarrow \overline{\Sigma}_{+}$ is onto; 
\item[iii)] $k_{-} >0$ everywhere on $S$.
 \end{itemize}
we shall call such an $S$ an {\em enclosing surface} in $\Sigma$. 
\end{definition} 

Let us briefly pause to explain the motivation behind the clauses $(i)-(iii)$ of this definition. First, it is meant as a convenient adaptation of Gannon's definition of a {\em regular near infinity} hypersurface, so item $(i)$ presents no novelty. Clause $(ii)$, however, might look somewhat opaque.  But it simply means that the (closure of the) outside of $S$ has only topological (or more precisely path-homotopic) complexities arising from having $S$ itself as a boundary. Specifically, {\em it means that every loop in the exterior of $S$ in $\Sigma$ is homotopic to a loop on $S$}. Note that this is certainly the case if $\overline{\Sigma}_{+} \equiv S \cup \Sigma_{+}$ is homeomorphic to $S \times [0, +\infty)$, as in the original Gannon-Lee theorem, but the condition as stated gives rise to the much wider set of topological possibilities which are likely to arise in higher dimensions. Finally, $(iii)$ says that $S$ is what Lee \cite{lee} designated as {\em uniformly convex}, and could also be called {\em inner trapped}. Physically, it is a condition which is naturally expected to hold if $S$ is thought of as a ''large surface in a weak gravitational field region''. Note that since it refers only to the inward-pointing familly of null geodesics, $S$ could independently be a (marginally) outer trapped surface as well. 

We are finally ready to state our main result.

\begin{theorem}
\label{maintheorem}
Let $(M,g)$ be an $n$-dimensional (with $n \geq 3$) null geodesically complete spacetime, which satisfies the null energy condition and admits an asymptotically regular hypersurface $\Sigma \subset M$. Suppose that $(M,g)$ is causally simple, $\Sigma$ admits a piercing, and $M$ is homeomorphic to $\Sigma \times \mathbb{R}$. Then, given an enclosing surface $S \subset \Sigma$, the group homomorphism $i_{\#}: \pi_1(S) \rightarrow \pi_1(\Sigma)$ induced by the inclusion $i: S \hookrightarrow \Sigma$ is surjective. In particular, if $S$ is simply connected, then so is $\Sigma$.  
\end{theorem}

The conclusion of the theorem can be more forcefully expressed in contrapositive terms: {\em if $\Sigma$ has some loop which is not homotopic to a loop entirely contained in $S$, then $M$ has at least one incomplete null geodesic}. We recover the Gannon-Lee conclusion in the particular case that $S$ is an $(n-2)$-sphere (for $n \geq 4$). 

The condition that $M$ is homeomorphic to $\Sigma \times \mathbb{R}$ can be interpreted as meaning that ``spatial topology does not change in time''. This is a technical, but probably not very restrictive condition, because of a classic theorem of Geroch \cite{geroch1} suggesting that spacetimes violating this condition would either violate causality or have singularities anyway. It is automatically satisfied if global hyperbolicity holds, by virtue of another well-known theorem by Geroch \cite{geroch2}. 

Recall that $(M,g)$ is {\em causally simple} if it is {\em causal} (i.e., admits no closed causal curves) and $J^{\pm}(p)$ are closed subsets of $M$ for all $p \in M$ \cite{HE,BE,MS}. In the next Section, we discuss this causal condition in more detail. For the moment, let us emphasize that if $\Sigma$ is an asymptotically regular {\em Cauchy} hypersurface of $(M,g)$ (so that $(M,g)$ is, in particular, globally hyperbolic), then the conditions that $(M,g)$ be causally simple, that $M \simeq \Sigma \times \mathbb{R}$, and that $\Sigma$ admits a piercing are automatically satisfied. 

A natural question is whether it is possible to weaken the conditions of causal simplicity and/or existence of a piercing in Theorem \ref{maintheorem}. Further investigation is necessary before we can settle this issue in a fully satisfactory way, but we can give a partial answer here, by introducing alternative assumptions for Theorem \ref{maintheorem} . In Ref. \cite{minguzzi}, its author obtains the striking result that {\em a chronological spacetime without past-directed null rays is globally hyperbolic} (the corresponding statement for future-directed null rays also being valid) \footnote{Recall that a spacetime $(M,g)$ is {\em chronological} if it admits no closed timelike curves, and a past-directed {\em null ray} is a past-inextendible causal curve $\gamma:[a,b) \rightarrow M$ with achronal image, which means that it can be parametrized as a null geodesic maximizing the Lorentzian arc-length between any two of its points. Future-directed null rays are defined time-dually.}. As pointed out in \cite{minguzzi}, it is possible to derive a singularity theorem from this result if we admit that some condition holds which guarantees that any past-complete null geodesic $\gamma:(-\infty,b] \rightarrow M$ admits a pair of conjugate points. A sufficient condition for this is, e.g, given in Refs. \cite{tipler, borde}:
\[
\lim_{s \rightarrow -\infty} [(b-s) \int_{-\infty}^s \mbox{Ric}(\gamma '(t),\gamma '(t))dt] >1.
\]
(A time-dual version of this condition can be given for future-complete null geodesics.) We shall call any such condition a {\em ray convergence condition} for past(future)-complete null geodesics. With the help of these results, we have: 

\begin{corollary}
\label{maincorollary}
Let $(M,g)$ be a chronological $n$-dimensional (with $n \geq 3$) null geodesically complete spacetime, which satisfies the null energy condition and a ray convergence condition for past-complete null geodesics. Suppose that there exists an asymptotically regular hypersurface $\Sigma \subset M$ admitting a piercing. Then, given an enclosing surface $S \subset \Sigma$, the group homomorphism $i_{\#}: \pi_1(S) \rightarrow \pi_1(\Sigma)$ induced by the inclusion $i: S \hookrightarrow \Sigma$ is surjective. 
\end{corollary}

{\em Proof:} From the chronology and the ray convergence condition for past-complete null geodesics, Theorem 10 of Ref. \cite{minguzzi} implies that $(M,g)$ is globally hyperbolic, and hence causally simple and homeomorphic to $\Sigma \times \mathbb{R}$. In the presence of the other conditions the result then follows by Theo. \ref{maintheorem}.
\qcd

Note that the condition that $\Sigma$ admits a piercing cannot be dropped in this Corollary, because even if $(M,g)$ is globally hyperbolic there is no a priori guarantee that $\Sigma$ is a Cauchy hypersurface. 

\section{Digression on Causal Simplicity}\label{causalsimplicity}


Since global hyperbolicity is a far more common assumption in many theorems of Lorentzian geometry, using causal simplicity as a hypothesis for our singularity theorem may seem somewhat contrived at first sight. Accordingly, the purpose of this Section is to review some of the implications of this causal condition and to argue that it is a natural condition to use in the current context.  

From the pioneering work of Penrose, Hawking, Geroch and others in the 1960's and 1970's to the present day, many specific causal features of spacetimes have been recognized, isolated and separately investigated. A partial (and somewhat rough) classification of spacetimes has been achieved in the form of the so-called {\em causal ladder} or {\em causal hierarchy} (see, for instance, \cite{MS} for a detailed description and recent results, and \cite{minguzzi} for a more complete version of the causal ladder). For definiteness, we reproduce it below.
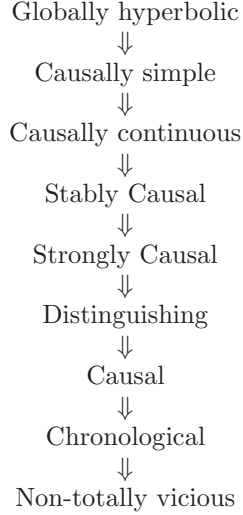
\[
\begin{array}{c}

\mbox{Globally hyperbolic}
\\
\Downarrow 
\\
\mbox{Causally simple}

\\
\Downarrow 
\\
\mbox{Causally continuous}

\\
\Downarrow 
\\
\mbox{Stably Causal}

\\
\Downarrow 
\\
\mbox{Strongly Causal}

\\
\Downarrow 
\\
\mbox{Distinguishing}

\\
\Downarrow 
\\
\mbox{Causal}

\\
\Downarrow 
\\
\mbox{Chronological}

\\
\Downarrow 
\\
\mbox{Non-totally vicious}

\end{array}
\]
In this ladder, each given condition for $(M,g)$ implies the ones below it and none of the implications can be reversed. As one can see, the condition of causal simplicity is second only to global hyperbolicity in strength. Is is usually defined by requiring that $(M,g)$ be {\em distinguishing} (i.e., $I^{+}(p) = I^{+}(q)$ or $I^{-}(p) = I^{-}(q)$ implies that $p=q$) together with the condition that $J^{\pm}(p)$ are closed subsets of $M$ for all $p \in M$ \cite{HE,BE}. However, as pointed out in \cite{MS}, in the presence of this second condition it is enough to require that $(M,g)$ is causal, and distinguishability follows. 

The condition that the causal pasts and futures of points in $M$ be closed can be presented in different but equivalent ways (see Lemma $3.67$ in \cite{MS}): 

\begin{proposition}
\label{closure} 
In {\em any} spacetime $(M,g)$, the following are equivalent:
\begin{itemize}
\item[i)]$J^{\pm}(p)$ are closed subsets of $M$ for all $p \in M$;
\item[ii)] If the sequences $(p_n)$ and $(q_n)$ in $M$ converging to the points $p$ and $q$, respectively, are such that $p_n \leq q_n$ for each $n \in \mathbb{N}$, then $p \leq q$;
\item[iii)] $J^{\pm}(K)$ are closed subsets of $M$ for every compact $K \subseteq M$.
\end{itemize}
\end{proposition}

Unlike causality or even strong causality, this condition is very delicate, in the sense that in general it will no longer hold if we delete points or closed sets from a spacetime $(M,g)$ in which it holds, as can easily be seen by deleting points of null cones in Minkowski spacetime. The reason is as follows. Let ${\cal U} \subseteq M$ be an open subset, $p \in {\cal U}$ such that $J^{+}(p)$ is closed, and let $(q_n)$ be a sequence of points in the causal future $J^{+}(p,{\cal U})$ converging to a point $q \in {\cal U}$. Since $J^{+}(p,{\cal U}) \subseteq J^{+}(p)$ and the latter set is closed, $q \in J^{+}(p)$. But in general $J^{+}(p,{\cal U}) \neq J^{+}(p)\cap {\cal U}$, so $q$ need not be in $J^{+}(p,{\cal U})$, and thus this set set may not be closed in ${\cal U}$. In this case, the spacetime $({\cal U},g|_{{\cal U}})$ will not be causally simple even if $(M,g)$ is. One exception occurs if ${\cal U}$ is {\em causally convex}, i.e., if any future-directed causal curve segment in $M$ with endpoints in ${\cal U}$ must actually be entirely contained in ${\cal U}$. In that case we {\em do} have $J^{+}(p,{\cal U})= J^{+}(p)\cap {\cal U}$. More important exceptions will be seen below.    

Now, recall that $(M,g)$ is globally hyperbolic if (1) it is strongly causal and (2) the ``causal diamonds'' $J^{+}(p) \cap J^{-}(q)$ are compact, for each $p,q \in M$. Since a causally simple spacetime is in particular strongly causal, it is precisely the second condition which must fail in a causally simple, non-globally hyperbolic spacetime. In physical terms, condition (2) means that no information which could be transmitted along causal curves can ``leak away" to infinity or to a naked singularity. Indeed, the whole point of the cosmic censorship conjecture is to argue that such a loss of predicability is unacceptable because it threatens physical predicabity, and then enforce global hyperbolicity in certain open subsets of spacetime (see, for instance, \cite{wald}, pp. 299-308, and references therein for the basic ideas) in order to prevent it. More specifically, the strongest version of the cosmic censorship conjecture is essentially the statement that a generic, physically realistic spacetime (in some suitable sense) must be globally hyperbolic \cite{wald}.

But even if one considers weaker forms of cosmic censorship, in which the requirement of global hyperbolicity is discarded, one might still end up with definite causal constraints on spacetime. For example, in Ref. \cite{BK}, the authors work with one of these weaker versions and show that it implies that spacetime is {\em pseudoconvex}, a condition closely related to causal simplicity (see \cite{BK} for details). We give below another example of how causal simplicity can arise from (a particular version of) the cosmic censorship requirement. 

In order to make our ideas precise, we need Penrose's notion of conformal infinity. Most of the following notions are rather standard, but we recall the general setting in order to fix some terminology. Let $f:M \rightarrow M'$ be a smooth embedding of $M$ as an open submanifold of the smooth manifold $M'$ with smooth boundary ${\cal J}$ such that $f(M) = M' \setminus {\cal J}$. On $M'$ we assume defined a smooth Lorentz metric $g'$ and a smooth function $\Omega : M' \rightarrow \mathbb{R}$ such that $f^{\ast} g' = (\Omega \circ f)^2 g$, $\Omega \circ f >0$ on $M$, and $\Omega(p) =0$ but $d\Omega_p \neq 0, \forall p \in {\cal J}$. Also, the time-orientation on $(M',g')$ is the unique one which is preserved by $f$. $(M',g')$ is what is called a {\em spacetime-with-boundary}, and we say that $(M,g)$ is {\em conformally embedded} therein. In this case, the {\em conformal boundary} (or {\em conformal infinity}) of $(M,g)$ is the boundary ${\cal J}$ of $M'$. Following the standard usage, we identify $M$ with $f(M)$ and omit explicit reference to the specific embedding $f$ from now on. In Ref. \cite{solisthesis}, some general causal properties of spacetimes-with-boundary are studied. A number of results which hold in spacetimes are no longer valid in spacetimes-with-boundary, so one must proceed with caution. A useful result (see the Appendix A of Ref. \cite{solisthesis} for a proof) is that every spacetime-with-boundary $(M',g')$ admits an extension to a spacetime (without boundary) of the same dimension. 

The particular version of cosmic censorship we shall use in our next result is adapted (but slightly different) from Ref. \cite{galloway3}, and presented as the conditions $(a)-(c)$ of Theorem \ref{causalitycool1} below. Accordingly, we shall say that a future-inextendible causal curve $\alpha$ is {\em visible} from a point $q$ in a spacetime (with or without boundary) if it is contained in $\overline{I^{-}(q)}$. Note that in this sense of the term `visible', $\alpha$ may lie entirely outside $J^{-}(q)$. One reason for this peculiar phrasing can be understood by picturing a future-inextendible geodesic generator $\gamma$ of the event horizon $\partial I^{-}({\cal J}^{+})\cap M$ in an asymptotically flat spacetime $(M,g)$ containing a black hole, where ${\cal J}^{+}$ denotes the future null infinity \footnote{In Prop.\label{causalitycool}, however, ${\cal J}^{+}$ can denote a different, more general type of future infinity.}. Suppose that we only regarded $\gamma$ as `visible' if it were in $J^{-}({\cal J}^{+})$. Then, if $\gamma$ were in $\partial I^{-}({\cal J}^{+})\cap M \setminus J^{-}({\cal J}^{+})$, then the condition (b) of Theorem \ref{causalitycool1} might not prevent $\gamma$ from being future-incomplete. But in that case, a slight perturbation of the metric, say, by ``opening up'' causal cones a bit, could make the incompleteness visible from future null infinity, in which case one would have a naked singularity (see \cite{galloway3,penroseCC} for additional motivation for this definition). 

With this notation, we have the following result:

\begin{proposition}
\label{causalitycool1}
Let $(M,g)$ be a causal spacetime conformally embedded in a spacetime-with-boundary $(M',g')$, with smooth boundary ${\cal J}$. Suppose that the following conditions hold:
\begin{itemize}
\item[(a)] ${\cal J}$ is the disjoint union of achronal submanifolds ${\cal J}^{+}$ and ${\cal J}^{-}$ of $M'$ and for every $p \in M$ there exist open neighborhoods ${\cal N}_{+}$ and ${\cal N}_{-}$ of ${\cal J}^{+}$ and ${\cal J}^{-}$, respectively, in $M'$, with $I^{\pm}(p,M) \cap {\cal N}_{\mp} = \emptyset$.  
\item[b)] Every future-inextendible null geodesic in $(M,g)$ visible from a point in ${\cal J}^{+}$ is future-complete.
\item[c)] Every future-complete null geodesic in $(M,g)$ has a future endpoint on ${\cal J}^{+}$.
\end{itemize}
Then, the domain of outer communication $D := I^{+}({\cal J}^{-},M') \cap I^{-}({\cal J}^{+},M')$ is causally simple (as an open submanifold of $M$ with the metric $g|_{D}$ and the induced time-orientation). 
\end{proposition}

{\em Proof:} First, we claim that $D_{\mp} := I^{\pm}({\cal J}^{\mp},M') \cap M$ are open subsets of $M$. Indeed, let $p \in D_{+}=I^{-}({\cal J}^{+},M') \cap M$, say, and let $\alpha:[0,1] \rightarrow M'$ be a past-directed timelike curve segment with $\alpha(0) \in {\cal J}^{+}$ and $\alpha(1) =p$. Since ${\cal J}^{+}$ and ${\cal J}^{-}$ are achronal, $\alpha$ does not intersect ${\cal J}^{+}$ in any other point and intersects ${\cal J}^{-}$ at most once. Suppose that for some $0<t_0<1$, $\alpha(t_0) \in {\cal J}^{-}$. Now, for some neighborhood ${\cal N}_{-}$ of ${\cal J}^{-}$ in $M'$, $I^{+}(p,M) \cap {\cal N}_{-} = \emptyset$, by condition $(a)$. However, by the continuity of $\alpha$, we can pick a number $\epsilon >0$ such that $t_0 + \epsilon <1$ and $\alpha([t_0, t_0 + \epsilon]) \subset {\cal N}_{-}$. But $\alpha(t_0 + \epsilon) \in I^{+}(p,M)$, a contradiction. Therefore, $\alpha$ does not intersect ${\cal J}^{-}$. Choose any $0<s_0 <1$. Since $p \in I^{-}(\alpha(s_0),M)$ and the latter set is open, there exists an open neighborhood $U \subseteq I^{-}(\alpha(s_0),M) \subseteq D_{+}$ of $p$. The case for $D_{-}$ is analogous, and the claim is established. 

Clearly, $D = D_{+} \cap D_{-}$ is thus open, and it is easy to check it is causally convex in $M$, so it is enough to show that $(D_{+}, g|_{D_{+}})$ is causally simple. Suppose not. Since causality holds in $(D_{+}, g|_{D_{+}})$, one can show (see, e.g., Prop. 1 of Ref. \cite{BK}), that there exist $p \in D_{+}$ and a maximal (i.e., achronal) null geodesic $\gamma:[0,b) \rightarrow D_{+}$ ($b \leq +\infty$), future-inextendible in $(D_{+},g|_{D_{+}})$, with $\gamma[0,b) \subset \partial_{D_{+}} I^{-}(p,D_{+})$.  It is easy to check that $I^{-}(p,D_{+}) = I^{-}(p,M)$, and $\partial_{D_{+}}I^{-}(p,D_{+}) \subseteq \partial_M I^{-}(p,M)$, so $\gamma[0,b) \subset \partial_{M} I^{-}(p,M)$. Therefore, if $\gamma$ had a future endpoint in $(M,g)$, this point would be $p$, which is not possible, so $\gamma$ is future-inextendible in $(M,g)$. 

Now, for some $q \in {\cal J}^{+}$, $I^{-}(p,M) \subseteq I^{-}(q,M')$, and hence $\gamma[0,b) \subset \overline{I^{-}(q,M')}$ (the closure is in $M'$). We conclude that $\gamma$ is visible from a point in ${\cal J}^{+}$, and by condition (b), it must be future-complete. However, by condition (a) there exists an open neighborhood ${\cal N}_{+}$ of ${\cal J}^{+}$ with ${\cal N}_{+} \cap I^{-}(p,M) = \emptyset$. Hence, ${\cal N}_{+} \cap \overline{I^{-}(p,M)} = \emptyset$ (here the closure is in $M$), which in turn means that $\gamma$ cannot have a future endpoint on ${\cal J}^{+}$, in contradiction with the condition (c). We conclude that $(D_{+}, g|_{D_{+}})$ is causally simple, and so is $(D, g|_{D})$, thus the proof is complete.
\qcd

The condition (a) in this Proposition is an analogue of the ``$i_0$-avoidance'' condition used in \cite{galloway3}, but here we do not impose either that $(M,g)$ be asymptotically flat or that any specific field equations like the Einstein equations hold therein, so we do not make any explicit reference to a spatial infinity $i_0$. Condition $(a)$ is meant, among other things, to avoid that the chronological past of a point in $M$ contains the whole ``future history'' of a material particle or photon. It is violated in spacetimes like Reissner-Nordstr\"{o}m (for points on the ``internal event horizon''), for example. However, this condition (and the conclusion) is clearly applicable to many situations in which ${\cal J}$ is spacelike (i.e., when $(M,g)$ is {\em asymptotically de Sitter}), or null (as in the asymptotically flat case). Moreover, in the proof we actually established that the ``region outside the black hole'', $I^{-}({\cal J}^{+},M') \cap M$, is causally simple, so the result also applies to some cosmological scenarios, where ${\cal J}^{-}$ may be empty. 

In spite of its large applicability, Prop. \ref{causalitycool1} still makes use of a reasonable but less familiar notion of cosmic censorship; one might want to adhere to the more standard versions which use global hyperbolicity (in which case causal simplicity holds trivially). While this can be done with relative success in asymptotically de Sitter and asymptotically flat spacetimes, in many (potentially) physically interesting cases, global hyperbolicity hopelessly fails, most notably in the (asymptotically) anti-de Sitter spacetimes studied in connection with the AdS/CFT and similar scenarios in string theories. In these cases, however, there is still the possibility that information apparently disappearing from spacetime can be retrieved by taking the conformal infinity into account. 

Following Refs. \cite{galloway4,solisthesis}, we define a {\em spacetime-with-timelike-boundary} to be a spacetime-with-boundary $(M',g')$ of dimension $n\geq 3$ whose smooth boundary ${\cal J}$ endowed with the induced metric is a Lorentzian manifold. A spacetime $(M,g)$ is {\em asymptotically anti-de Sitter} if it can be conformally embedded in a spacetime-with-timelike-boundary of the same dimension. Spacetimes-with-timelike-boundaries can also be studied \cite{solisthesis} in contexts where ``timelike tubes'' or ``finite infinity'' are relevant (see for instance \cite{galloway2}). Just as in the case without boundary, we say that a spacetime-with-boundary $(M',g')$ is {\em globally hyperbolic} if it is strongly causal and $J^{+}(p,M') \cap J^{-}(q,M')$ is compact, for each $p,q \in M'$ \cite{galloway4}. 

In this setting, we have the following result: 

\begin{proposition}
\label{causalitycool2}
Let $(M,g)$ be an asymptotically anti-de Sitter spacetime, conformally embedded in a spacetime-with-timelike-boundary $(M',g')$ with conformal boundary ${\cal J}$. Suppose that for every $p \in M$, $J^{+}(p,M')\cap M \subseteq J^{+}(p,M)$. In this case, if $(M',g')$ is globally hyperbolic, then $(M,g)$ is causally simple. 
\end{proposition}

{\em Proof.} $(M',g')$ being strongly causal means that $(M,g)$ is also (strongly) causal. Let $p \in M$, and let $(q_n)$ be sequence in $J^{-}(p,M)$ converging to some $q \in M$. By the Prop. 3.17 of Ref. \cite{solisthesis}, the sets $J^{\pm}(p,M')$ are closed (in $M'$), so $q \in J^{-}(p,M')\cap M$, i.e., $p \in J^{+}(q,M') \cap M$, and by our assumption, $p \in J^{+}(q,M)$, i.e., $q \in J^{-}(p,M)$. Thus $J^{-}(p,M)$ and (by an entirely analogous argument) $J^{+}(p,M)$ are closed, so we conclude that $(M,g)$ is causally simple.
\qcd 
  
The condition of global hyperbolicity on $(M',g')$ in this Proposition is of course meant to capture the idea of ``retrieving information at infinity''. The other assumption, namely that for every $p \in M$, $J^{+}(p,M')\cap M \subseteq J^{+}(p,M)$, simply means that, given $p$ and $q$ points in $M$ and $\alpha$ a future-directed causal curve in $M'$ from $p$ to $q$, it is always possible to find a future-directed causal curve $\beta$ from $p$ to $q$ entirely contained in $M$ (i.e., $\beta$ avoids the conformal boundary). Geometrically, it is an extrinsic constraint on the boundary ${\cal J}$ positing that causal futures in $M'$ should not be ``squashed'' against ${\cal J}$. This condition is natural enough if one has the standard conformal embeddings of spacetimes like anti-de Sitter and Schwarzschild-anti-de Sitter in mind, and indeed it is hard to imagine how it can be violated in any spacetime $(M,g)$ satisfying, say, the Einstein vacuum equation with negative cosmological constant in a neighborhood of conformal infinity, except by artificial constructions. Although the condition is stated in a conformally invariant manner, which is preferable from the standpoint of causal theory, it would be interesting to derive it from more physically motivated conditions. We shall not attempt this here. 
 
\section{Proof of Theorem \ref{maintheorem}}\label{detailedproofs}

We first prove the following auxiliary result:

\begin{proposition}
\label{auxiliary}
Let $(M,g)$ be an $n$-dimensional, causally simple, null geodesically complete spacetime (with $n \geq 3$) which satisfies the null energy condition and admits an asymptotically regular hypersurface $\Sigma \subset M$. Suppose that $\Sigma$ admits a piercing. Then, given an enclosing surface $S \subset \Sigma$, the closure of the inside of $S$, $\overline{\Sigma}_{-} \equiv S \cup \Sigma_{-}$, is compact. 
\end{proposition}

{\em Proof.} The proof follows the general idea set forth in \cite{lee}, Theorem 1, adapting some standard arguments in the proofs of the Penrose-Hawking singularity theorems (see, for instance, \cite{oneill}, Prop.14.60). 

 Since $S$ is compact, there exists a number $k_0 >0$ for which $k_{-}(p) \geq k_0, \forall p \in S$. Put $b=\frac{1}{k_0}$ and define 
\[
B= \{ t\cdot K_{-}(p)  \in NS \,:\,p \in S \mbox{ and } 0 \leq t \leq b \}, 
\]
where $NS$ denotes the normal bundle of $S$ in $M$. Since $B$ is the image of $S \times [0,b]$ under the continuous map $f:S \times \mathbb{R} \rightarrow NS$ given by $f(p,t) = t \cdot K_{-}(p), \forall t \in \mathbb{R}, \forall p \in S$, $B$ is compact. From the null geodesic completeness, $B$ is contained in the domain of the normal exponential map 
$\exp_{\bot}$, and $\exp_{\bot}(B) \subseteq M$ is compact. 

Now let $T:= \partial I^{+}\left( \Sigma_{+} \right) \setminus \left( \Sigma_{+} \right)$. The remaining part of the proof will be presented as a series of claims. 

\vspace{.2cm}

{\em Claim 1}: $T \subseteq E^{+}(S) := J^{+}(S) \setminus I^{+}(S)$.  

\vspace{.2cm}

The strategy here is to show that $T \subseteq \partial I^{+}(S)$. Since $S$ is compact, $J^{+}(S)$ is closed by our assumption of causal simplicity (see Proposition \ref{closure}), and we must have $E^{+}(S) = \partial J^{+}(S) = \partial I^{+}(S)$. Thus, let $p \in T$. If $p \in \overline{\Sigma}_{+} \setminus \left( \Sigma_{+}\right)=\partial_{\Sigma} \left( \Sigma_{+}\right) = \partial_{\Sigma} \Sigma_{-} \equiv S \subseteq E^{+}(S)$, we are done. Otherwise, since $p \notin  \overline{\Sigma}_{+}$, given a ${\cal U} \subseteq M$ open neighborhood of $p$, we can assume without loss of generality that it does not intersect $\overline{\Sigma}_{+}$. Pick any $p_{\pm} \in I^{\pm}(p) \cap {\cal U}$. Since $\partial I^{+}\left( \Sigma_{+} \right)$ is an achronal boundary, $p_{-}$ is not in $I^{+}\left( \Sigma_{+} \right) = I^{+}\left(\overline{\Sigma}_{+}\right)$, and in particular not in $I^{+}(S)$, while $p_{+} \in I^{+}\left( \Sigma_{+} \right)$. In particular, $C := I^{-}(p_{+}) \cap \Sigma$ is not empty, and intersects $\Sigma_{+}$. Moreover, $p \in I^{+}(\Sigma)$, and shrinking ${\cal U}$ if necessary we can assume ${\cal U} \subseteq I^{+}(\Sigma)$, thus in particular $p_{-} \in I^{+}(\overline{\Sigma}_{-})$. Since $p_{-} <<p<<p_{+}$, $C$ must also intersect $\overline{\Sigma}_{-}$. 

Now, let $X:M \rightarrow TM$ be a piercing for $\Sigma$. The maximally extended integral lines of $X$ emanating from points of $M$ will reach $\Sigma$, thus inducing an open (i.e., maps open sets onto open sets), continuous map $\rho_{X}: M \rightarrow \Sigma$ (see, e.g., \cite{oneill}, Prop. 14.31) onto $\Sigma$ which leaves $\Sigma$ pointwise fixed. 

Let $q,q' \in C$. Since $I^{-}(p_{+})$ is connected, we can find a continuous curve $\alpha: [0,1] \rightarrow I^{-}(p_{+}) \cap J^{+}(\Sigma)$ connecting $q$ and $q'$. The curve $\rho_{X} \circ \alpha$ is thus a continuous curve in $C$ connecting $q$ and $q'$, and we conclude that $C$ is connected. But then $C \cap S \neq \emptyset$, since $S$ is the common boundary in $\Sigma$ of  $\Sigma_{+}$ and $\overline{\Sigma}_{-}$. We conclude that $p_{+} \in I^{+}(S)$, and thus that ${\cal U}$ intersects both $I^{+}(S)$ and $M \setminus I^{+}(S)$, which establishes that $p \in \partial I^{+}(S)$.   

\vspace{.2cm}

{\em Claim 2}: $T$ is compact. 

\vspace{.2cm}

The idea behind the proof of this claim is to show that $T \subseteq \exp_{\bot} (B)$; since $T$ is clearly closed, the result will follow. Given $p \in T$, either $p \in \overline{\Sigma}_{+} \setminus \left( \Sigma_{+}\right)=\partial_{\Sigma} \left( \Sigma_{+}\right) = \partial_{\Sigma} \overline{\Sigma}_{-} \equiv S \subseteq \exp_{\bot} (B)$, or else by Claim 1 there exists a null, future-directed geodesic segment $\g: [0,1] \rightarrow M$ with image in $\partial I^{+}\left( \Sigma_{+} \right)$, with $q=\g(0) \in S$, $\g(1) = p$, and without focal points before $p$.  

Now, $v =\g '(0) \in NS$ is a future-directed null vector, and we must have $v=s \cdot K_{\pm}(q)$ for some number $s >0$. It is not difficult to check that if $v$ were parallel to $K_{+}(q)$, one would have $p \in I^{+}\left( \Sigma_{+} \right)$, again a contradiction; thus, we conclude that $v= s \cdot K_{-}(q)$. Therefore, the absence of focal points before $p$ means, using the null energy condition (see, e.g., Prop. 10.43 of \cite{oneill}), that
\[
1 \leq \frac{1}{s \cdot k_{-}(q)} \leq \frac{1}{sk_0} \equiv \frac{b}{s},
\]
but then $0 <s \leq b$, and hence $v \in B$. Since $p = \g(1) \equiv \exp_{\bot} (v)$, the claim is proved. 

\vspace{.2cm}

{\em Claim 3}: $\rho_{X}(T)= \overline{\Sigma}_{-}$. 

\vspace{.2cm}

Clearly $\rho_{X}(T)\subseteq \overline{\Sigma}_{-}$, and $S \subseteq T$, so $S = \rho_{X}(S) \subseteq \rho_{X}(T)$. Thus $\overline{\Sigma}_{-} \setminus \rho_{X}(T) \subseteq \Sigma_{-}$. If the claim is false, then $\overline{\Sigma}_{-} \setminus \rho_{X}(T) \neq \emptyset$, and we must have $\partial_{\Sigma}\rho_{X}(T) \cap \Sigma_{-} \neq \emptyset$. Hence we can pick $p \in \partial_{\Sigma}\rho_{X}(T) \cap \Sigma_{-}$. Since $\rho_{X}(T)$ is compact, $p = \rho_{X}(q)$ for some $q \in T$. If $q\in \overline{\Sigma}_{+} \setminus \left(\Sigma_{+} \right) \equiv S$, then $\rho_{X}(q)=q=p \in \Sigma_{-}$, a contradiction. Therefore, we can assume that $ q \in \partial I^{+}\left( \overline{\Sigma}_{+} \right) \setminus \left( \overline{\Sigma}_{+} \right)$. Since the latter set is a topological hypersurface, we can choose a neighborhood ${\cal V}_0$ of $q$ in $M$ with ${\cal V}_0 \cap \Sigma = \emptyset$, ${\cal V}_0 \cap T$ open in  $\partial I^{+}\left( \overline{\Sigma}_{+} \right) \setminus \left( \overline{\Sigma}_{+} \right)$ and $\rho_{X}({\cal V}_0) \subseteq \Sigma_{-}$. 

Now, let $ \Psi: {\cal U}_0 \times (-\epsilon,\epsilon) \rightarrow M$ be a local flow of $X$ with $\mbox{Im} \Psi \subseteq {\cal V}_0$ and $q \in {\cal U}_0$. Put $\Psi_0 := \Psi|_{\left({\cal U}_0 \cap T\right) \times (-\epsilon,\epsilon)}$ and ${\cal W} := \mbox{Im} \Psi_0$. $\Psi_0$ is clearly one-to-one, since $T$ is achronal; therefore, by Invariance of Domain, ${\cal W}$ is open in $M$ and $\Psi_0$ is a homeomorphism onto ${\cal W}$. But then $p \in \rho_{X}\left({\cal U}_0 \cap T\right) = \rho_{X}({\cal W})$, and since the latter set is open in $\Sigma$ because $\rho_{X}$ is open, we conclude that $p$ is in the $\Sigma$-interior of $\rho_{X}(T)$, in contradiction with the fact that $p$ must be in the $\Sigma$-boundary of $\rho_{X}(T)$. The claim now follows, and the proof is complete.  
\qcd

{\em Proof of Theorem \ref{maintheorem}}:
\newline
Let $\phi : \tilde{M} \rightarrow M$ be a connected smooth covering of $M$ such that $\phi_{\#}(\pi_1(\tilde{M})) = j_{\#} \left(\pi_1(S) \right)$, where $j:S \hookrightarrow M $ is the inclusion in $M$ of $S$. We endow $\tilde{M}$ with the pullback metric $\tilde{g} = \phi^{\ast}g$ and induced time-orientation, so that $(\tilde{M},\tilde{g})$ is a spacetime, locally isometric  to $(M,g)$. Since $M$ is homeomorphic to $\Sigma \times \mathbb{R}$, the inclusion map $m: \Sigma \hookrightarrow M$ induces an isomorphism $m_{\#}: \pi_1(\Sigma) \rightarrow \pi_1(M)$. In particular, $\tilde{\Sigma} := \phi^{-1}(\Sigma)$ is connected. The restriction $\phi_{\Sigma}: \tilde{\Sigma} \rightarrow \Sigma$ of the map $\phi$ is itself a (Riemannian) covering. We claim that this covering is actually trivial, i.e., a diffeomorphism. Suppose for the moment that this is indeed the case, and denote by $\tilde{m}: \tilde{\Sigma} \hookrightarrow \tilde{M}$ and $i: S \hookrightarrow \Sigma$ the respective inclusions. Let $y \in \pi_1(\Sigma)$. We have $m_{\#}(y) \equiv (\phi \circ \tilde{m} \circ \phi_{\Sigma}^{-1})_{\#}(y) =\phi_{\#}(\tilde{m} \circ \phi_{\Sigma}^{-1})_{\#}(y) \in \phi_{\#}(\pi_1(\tilde{M})) = j_{\#} \left(\pi_1(S) \right)$, and hence there exists $x \in \pi_1(S)$ such that $m_{\#}(y) = j_{\#}(x) \equiv (m \circ i)_{\#}(x) = m_{\#} (i_{\#}(x))$, whence we conclude that $y = i_{\#}(x)$, which proves the theorem.   

Let us then establish the claim. First let us fix a local deformation of $S$, $F:S \times (-1,1) \rightarrow \Sigma$, and let $V:= {\cal U}_F^{-}\cup S \cup \Sigma_{+}$. Then, $\overline{\Sigma}_{+}$ is a deformation retract of $V$, and therefore $\pi_1(\overline{\Sigma}_{+})$ and $\pi_1(V)$ are isomorphic. We show that for each connected component $\tilde{V}$ of $\phi_{\Sigma}^{-1} (V)$, the restriction $\phi_{V}: \tilde{V} \rightarrow V$ of $\phi$ is a diffeomorphism. Indeed, $\phi_{V}$ is a smooth covering, and hence a local diffeomorphism, so we only need to show it is one-to-one. Given $\tilde{p},\tilde{q} \in \tilde{V}$ with $\phi_{V} (\tilde{q}) = \phi_{V} (\tilde{p}) = p \in V$, we can pick a path $\tilde{\alpha}:[0,1] \rightarrow \tilde{V}$ from $\tilde{p}$ to $\tilde{q}$, so that $\phi \circ \tilde{\alpha}$ is a loop in $V$, which by our assumption about $\pi_1(S)$ and $\pi_1(V) \simeq \pi_1(\overline{\Sigma}_{+})$ in Def. \ref{asympticallyregular}, item (ii), is homotopic to a loop in $S$. Since $\phi_{\#}(\pi_1(\tilde{M})) = j_{\#} \left(\pi_1(S) \right)$ we can find a loop $\tilde{\beta}$ in $\tilde{M}$ fixed-endpoint homotopic to $\tilde{\alpha}$, which in turn means that we must have $\tilde{p}=\tilde{q}$, establishing the injectivity of $\phi_{V}$. Now, suppose $\phi_{\Sigma}$ is not a diffeomorphism. Then, since $S \subset V$ and $\phi_{V}$ is a trivial covering, each connected component of $\phi_{\Sigma}^{-1}(S)$ is a diffeomorphic copy of $S$ itself, and there is more than one such component. Moreover, each one of these components also separates $\tilde{\Sigma}$. Let $\tilde{S}_1$ and $\tilde{S}_2$ be any two of these copies of $S$, contained respectively in the components $\tilde{V}_{1}$ and $\tilde{V}_{2}$ of $\phi^{-1}_{\Sigma}(V)$, which are disjoint copies of $V$ contained in $\tilde{\Sigma}$. We have a diffeomorphic copy of the closed non-compact set $\overline{\Sigma}_{+} \subseteq V$ in each connected component of $\phi_{\Sigma}^{-1} (V)$. Denote by $\tilde{C}_{i}$ the copy of $\overline{\Sigma}_{+}$ contained in $\tilde{V}_{i}$ ($i=1,2$).  Since $\tilde{\Sigma}$ is connected and is separated by $\tilde{S}_{1}$, the copy $\tilde{C}_{2}$ containing $\tilde{S}_2$ must be contained in the set $\tilde{S}_1 \cup \tilde{\Sigma}^{(1)}_{-}$, where $\tilde{\Sigma}^{(1)}_{-} := \tilde{\Sigma} \setminus \tilde{C}_{1}$, for otherwise $\tilde{V}_{1}$ would intersect $\tilde{V}_{2}$. However, it is not difficult to see that the hypotheses of Proposition \ref{auxiliary} will hold for $\tilde{\Sigma}$, $(\tilde{M}, \tilde{g})$ and $\tilde{S}_1$, and hence $\tilde{S}_1 \cup \tilde{\Sigma}^{(1)}_{-}$ must be compact. This contradiction establishes the claim, so the proof is complete. 
\qcd 

\section{Concluding Remarks}\label{conclusion}
There are a few more points about the scope of our result that are worth mentioning:
\begin{itemize}
\item[a)] The null convergence condition used here can of course be interpreted as arising from a corresponding condition on the energy-momentum tensor when suitable field equations are assumed. As usual in singularity theorems, it is used to ensure that certain null geodesics emanating from an enclosing surface $S\subset \Sigma$ have focal points. It can be weakened to an averaged null energy condition on future-inextendible geodesics starting at $S$, or on (past and future) inextendible null geodesics provided it is supplemented by the generic condition \cite{borde}. 
\item[b)] The condition of causality we impose, and even of chronology, is dispensable in our proof as long as the causal futures and pasts of points remain closed, and our piercing condition continues to hold. Our choice of a stronger causal condition is due to its seemingly greater physical applicability. 
\item[c)] As for the question of whether one can omit or change the assumptions of causal simplicity and/or existence of a piercing in our main theorem, a partial answer is given via Corollary \ref{maincorollary}. Its lesson is that we can at least weaken the causal simplicity assumption down to chronology provided we supplement the null energy condition with a convergence condition on null rays. In this Corollary, the existence of a piercing still remains as a fundamental premiss, but it {\em might} be an artifact of the particular arguments we gave, and could perhaps be dispensed with in an alternative proof. 
\item[c)] Although Gannon \cite{gannon1} offered a version of his theorem with weaker causal assumptions than global hyperbolicity (see \cite{gannon1}, Theo. 2.2), his proof of this part is based on a flawed assumption (cf. \cite{gannon1}, pg. 2366, second paragraph), as pointed out by Galloway in Ref. \cite{gallowayagain}, and so global hyperbolicity remains as the underlying causal condition of the original Gannon-Lee theorem. The causal simplicity condition in our version of the theorem, on the other hand, yields applications in much wider contexts, which include certain asymptotically anti-de Sitter spacetimes. 
\end{itemize} 

There has been tremendous progress in mathematical Relativity and Lorentzian Geometry since the appearance of the Gannon-Lee theorem, but singularity theorems of this sort retain their importance, both from a purely geometric standpoint and for our understanding of gravity. The generalization presented here would be nearly impossible to anticipate without the background of recent insights. This result suggests that we look at the Gannon-Lee theorem not as an {\em absolute} restriction on the spatial topology, which may not be natural in certain contexts, but rather as a {\em relation} between spatial topology and the topology of certain embedded surfaces. Regarded in this new light, this old theorem may ``come back to life" and hopefully be a source of new ideas. 
\newpage  

%
\end{document}